\documentstyle[12pt,a4,psfig,here]{article}
\newcommand{\be}{\begin{equation}}
\newcommand{\ee}{\end{equation}}
\newcommand{\bea}{\begin{eqnarray}}
\newcommand{\eea}{\end{eqnarray}}
\newcommand{\bean}{\begin{eqnarray*}}
\newcommand{\eean}{\end{eqnarray*}}

\begin{document}
\begin{flushright}
HUPD-9719\\
hep-ph/9711260
\end{flushright}

\begin{center}
{\LARGE\bf QCD Higher Order Corrections\\
             to $g_1 (x)$ at Small $x$}
    {}\footnote{Talk presented by J. Kodaira. To appear in:
     Proceedings of the Workshop on Deep Inelastic Scattering
     off Polarized Targets: Theory Meets Experiment,
      DESY Zeuthen, September 1-5, 1997.}

\vspace{1cm}
{Yuichiro KIYO, Jiro KODAIRA and Hiroshi TOCHIMURA}

\vspace*{1cm}
{\it Dept. of Physics, Hiroshima University Higashi-Hiroshima
      739, JAPAN}\\

\vspace*{2cm}

\end{center}

\begin{abstract}
The small $x$ behavior of the flavor non-singlet
$g_{1}$ structure function is analysed numerically by
taking into account the
all-order resummation of $\alpha_{s} \ln^{2}x $ terms.
We include a part of the next-to-leading logarithmic corrections
coming from the resummed \lq\lq coefficient function\rq\rq\ which
are not considered in the literatures to respect
the factorization scheme independence.
The resummed coefficient function turns out to
give unexpectedly large suppression effects over the
experimentally accessible range of $x$ and $Q^{2}$.
This fact implies that the higher order logarithmic
corrections are very important for $g_{1}$ in the small $x$ region.
\end{abstract}

\section{Introduction}

\vspace{1mm}
\noindent
The behavior of the structure function in the small
Bjorken $x$ region has recently received much attention of the
physicists~\cite{intro}.
This region corresponds to the so-called Regge limit.
So we naively expect that the soft physics (Regge theory) may
explain the small $x$ behavior of the structure function.
However the steep rise of the structure function in this
region observed by the HERA experiments contradicts with this naive
expectation. The physics at small $x$ is now one of the most interesting
subjects and many people believe that this problem could be
handled in the context of the QCD perturbation theory~\cite{bfkl}.

In the case of the polarized structure function $g_{1}$,
we have not yet had data at very small $x$~\cite{smc}.
However it is very important and desirable to know
the small $x$ behavior of $g_1$ in the light of the Bjorken
and Ellis-Jaffe sum rules. The naive Regge theory tells us,
\[ g_{1} \sim x^{\alpha}  , \qquad  0 \leq \alpha \leq 0.5 \ ,\]
at small $x$ and $Q^2$ and this implies a \lq\lq flat\rq\rq\
input parton density. Although the recent detailed
analyses~\cite{grsv} of the existing data in the DGLAP
approach do not necessarily require
an input function which contradicts with the naive Regge
prediction, it is inevitable to take into account the
effects from the  $\ln x $ corrections, which appear in the
perturbative calculations, to get more precise predictions
on the small $x$ behavior of $g_1$~\cite{talk1}. 

Bartels, Ermolaev and Ryskin~\cite{bartels} have given
the resummed expression for the $g_{1}$ structure function by using
the Infra-Red Evolution Equation and confirmed the old
result by Kirschner and Lipatov~\cite{kili}.
They claim that the resummation effects may be very important.
On the other hand, the numerical analysis by Bl\"umlein and
Vogt~\cite{blvo} shows
that there are no significant contributions from the resummation
of the leading logarithmic (LL) corrections in the HERA kinematical
region ( $x\sim 10^{-3}$). This controversial aspect might
come from the fact that the resummed \lq\lq coefficient function\rq\rq \ is
considered in Ref.~\cite{bartels} but not in Ref.~\cite{blvo}. 
Bl\"umlein and Vogt did not include the
resummed coefficient function because it
falls in the next-to-leading logarithmic (NLL) corrections
and depends on the factorization scheme adopted. 
It is also to be noted that the evolution, in general, strongly
depends on the input parton densities.
If one chooses a steep input function, the perturbative
contribution will be completely washed away.
So it will be interesting to see also the sensitivity of the
results to the choice of the input densities.

In this report, we discuss the numerical impact  
of the all-order $\ln x$ resummation on the small $x$
behavior of $g_{1}$ (non-singlet part).
We consider two different input densities: one is flat
corresponding to the naive Regge prediction and
the other is steep in the small $x$ region.
The coefficient function can not
be included consistently at present since the anomalous
dimension has been calculated only at the LL order.
However we consider the effects of the coefficient
function because we could firstly clarify the
above controversial aspect and secondly get some
idea about the magnitude of the NLL order
corrections in the resummation approach. Details of the
calculations may be found in ref.~\cite{kkt}. 

%---------- Section 2 ----------------------------------------
\section{Resummation of $\ln x$ terms } 

\vspace{1mm}
\noindent
The flavor non-singlet part of $g_{1}$ in the moment space is given by,
\[ g_{1}(Q^{2},N) \equiv \int^{1}_{0}dx x^{N-1} g_{1}(Q^{2},x)
  = \frac{\langle e^{2} \rangle}{2}
     C (\alpha_{s}(Q^{2}),N) \Delta q (Q^{2},N)\ ,\]
where
\[ C (\alpha_{s}(Q^{2}),N) \equiv \int^{1}_{0}x^{N-1}
    C (\alpha_s (Q^{2}),x) \ \ ,\ \ \Delta q (Q^{2},N)
     \equiv \int^{1}_{0}x^{N-1} \Delta q (Q^{2},x)\ ,\]
and $\Delta q (Q^{2},x)$ ($C$) is the flavor non-singlet
combination of the polarized parton densities (the coefficient function).
$\langle e^{2} \rangle$ is the average of quark's electric charge. 
The DGLAP equation reads,
\be
  Q^{2} \frac{\partial}{\partial Q^{2}} \Delta q (Q^{2},N) 
   = - \gamma (\alpha_{s}(Q^{2}),N)
   \Delta q (Q^{2},N) \ .\label{eqn:rgeq} 
\ee
Here the anomalous dimension $\gamma$ is the moment of the
\lq\lq splitting\rq\rq\ function.

The coefficient function $C (\alpha_{s},N)$ and the
anomalous dimension $\gamma (\alpha_{s},N)$ may be expanded
in the powers of $\alpha_s$,
\[ C (\alpha_{s},N) = 1 + \sum_{k=1}^{\infty}
     c^{k}(N) \bar{\alpha}_{s}^{k} \ \ ,\ \ \gamma (\alpha_{s},N)
   = \sum_{k=1}^{\infty} \gamma^{k}(N) \bar{\alpha}_{s}^{k} \ .\]
where $\bar{\alpha}_{s} \equiv \frac{\alpha_s}{4\pi}$ .
The singular behaviors of these
functions as $x \to 0$ appear as the pole singularities at $N=0$.
When $x$ is finite, it may be enough to compute them
to the fixed-order of perturbation. 
In the small $x$ region, however, the fixed-order calculation
becomes questionable since there appear $\ln ^n x$ type corrections.
If these $\ln ^n x$ terms
compensate the smallness of the coupling constant $\alpha_s$,
we must resum the perturbative series to the all orders to
get a reliable prediction. The explicit next-to-leading order (NLO)
calculations~\cite{koda}~\cite{van} in the $\overline{{\rm MS}}$ scheme
show a strong singularity at $N=0$,
\bea
 c^{1}(N) &=& 2 C_F \frac{1}{N^2} + {\cal O} \left( \frac{1}{N}
     \right)\ , \nonumber\\
 \gamma^{2}(N) &=& 4 (3 C_F^2 - 2 C_A C_F ) \frac{1}{N^3}
      + {\cal O} \left( \frac{1}{N^2} \right)\ . \label{NLO}
\eea
These strong singularities (double logarithmic corrections)
will persist to all orders of perturbative series.
Indeed, at the $k$-th loop, the anomalous dimension and the coefficient
function are expected to behave as,
\be
 \gamma^{k}(N) \sim N \left( \frac{1}{N^2} \right)^k \quad , \quad
   c^{k} (N) \sim \left( \frac{1}{N^2} \right)^k \ .\label{expect}
\ee
Our task is to resum these terms to all-orders in the perturbative
expansion.

The resummation of $\ln x$ singularities for $g_1$
has been done in Refs.~\cite{bartels}~\cite{kili}.
The result for the \lq\lq parton (quark) \rq\rq\ target with
the fixed $\alpha_s$\footnote{In the genuine LL approximation,
the strong coupling constant should be taken as a fixed parameter.}
reads,
\[
 g_{1}^{\rm parton} (x , Q^2 )= \frac{e_i^2}{2}
   \int_{ c-i \infty}^{c+i \infty} \frac{d N}{2 \pi i} x^{- N}
   \left( \frac{Q^2}{\mu^2} \right)^{f_0^- (N) / 8\pi^2}
   \frac{N}{N - f_0^- (N) / 8\pi^2} \ ,
%\label{partonresum}
\]
where $\mu$ is an arbitrary mass scale which regularizes the
infrared and/or mass singularities. From this expression we could
identify the resummed anomalous dimension $\hat{\gamma}$ and the
coefficient function $\hat{C}$ to be,
\bea
   \hat{\gamma}(\alpha_{s},N) 
    &\equiv& \lim_{N \to 0} \gamma(\alpha_{s},N) = -
   f^{-}_{0}(N)/8 \pi^{2} \ ,\label{resuma}\\
  \hat{C}(\alpha_{s},N) &\equiv& \lim_{N \to 0} C(\alpha_{s},N)
     = \frac{N}{N-f^{(-)}_{0}(N) / 8\pi^{2}} \ .\label{resumc}
\eea
Here $f_{0}^{-}$ is given by,
\[   f^{-}_{0}(N) = 4 \pi^{2}N
  \left( 1-\sqrt{1- 8 C_F \frac{\bar{\alpha}_s}{N^{2}}
  \left[ 1-\frac{1}{2\pi^{2}N}f^{(+)}_{8}(N) \right]} \right) \ .\]
with
\[ f^{+}_{8} (N) = 16 \pi^2 N_{c} \bar{\alpha}_s \frac{d}{dN}
   \ln(e^{z^{2}/4}D_{-1/2N_{c}^{2}}(z)) \quad , \quad
           z = \frac{N}{\sqrt{2 N_{c} \bar{\alpha}_s}}\ .\]
$ D_{p}(z)$ is the parabolic cylinder function.

Now it will be instructive to re-expand Eqs.(\ref{resuma},\ref{resumc})
in terms of $\alpha_s$ to see whether these formulae sum up the
most singular terms of the perturbative series.
The expressions expanded up to ${\cal O} (\alpha_s^4 )$ read,
\bea
  \hat{\gamma} &=& - N \left[
     2 C_{F} \left( \frac{\bar{\alpha}_{s}}{N^{2}} \right)
   + 4 C_{F} \left(C_{F}+\frac{2}{N_{c}} \right)
     \left( \frac{\bar{\alpha}_{s}}{N^{2}} \right)^{2} \right.
            \nonumber \\
   &+& \left. 16 C_F \left( C_{F}^2 + 2 \frac{C_{F}}{N_{c}}
       - \frac{1}{2N_{c}^{2}} - 1 \right)
    \left( \frac{\bar{\alpha}_{s}}{N^{2}} \right)^{3} \right] + \,
    \cdots = N \sum^{\infty}_{k=1} \hat{\gamma}^{k}
   \left( \frac{\bar{\alpha}_{s}}{N^{2}} \right)^{k} ,\label{ptg}\\
 \hat{C} &=& 1 + 2C_{F} \left( \frac{\bar{\alpha}_{s}}{N^{2}} \right)
     + 8 C_F \left( C_{F} + \frac{1}{N_{c}} \right)
   \left( \frac{\bar{\alpha}_{s}}{N^{2}} \right)^{2} \nonumber \\
   &+& 8 C_F \left( 5 C_{F}^{2} + 8 \frac{C_F}{N_c}
        - \frac{1}{N_c^2} - 2 \right)
     \left( \frac{\bar{\alpha}_{s}}{N^{2}} \right)^{3}
    + \, \cdots = \sum^{\infty}_{k=0} \hat{c}^{k}
   \left( \frac{\bar{\alpha}_{s}}{N^{2}} \right)^{k} . \label{ptc}
\eea
These results coincide with the previous expectation of
Eq.(\ref{expect}). Furthermore one can easily see that the resummed expressions
Eqs.(\ref{resuma},\ref{resumc}) reproduce
the known NLO results Eqs.(\ref{NLO}) in the $\overline{\rm MS}$
scheme~\footnote{This statement seems questionable beyond the NLO
as stressed by J. Bl\"umlein.}. 
Therefore, it is quite plausible that
Eqs.(\ref{resuma},\ref{resumc}) correctly sum up
the \lq\lq leading\rq\rq\ singularities to all orders.

Here a comment is in order concerning the scheme dependence.
The anomalous dimension and the coefficient
function individually depend on the factorization scheme.
(Unfortunately we do not have by now any appropriate factorization
theorems to the problem discussed in this report.)
In particular, the resummed \lq\lq coefficient function\rq\rq\ does
not have any physical meaning until the scheme dependent part
of the anomalous dimension is calculated in the same scheme.
To clarify this issue, let us write
the above results in the form which corresponds
to the so-called DIS scheme.
The parton densities and anomalous dimension in the DIS scheme
are obtained by making the transformations,
\[ \Delta q
 \rightarrow  \Delta q^{DIS} \equiv C \Delta q \ \ ,\ \ \gamma^{DIS}
     \equiv C \gamma C^{-1} - \beta(\alpha_{s})
    \frac{\partial}{\partial \alpha_{s}} \ln C \ .\]
Using the resummed $\hat{\gamma}$ and $\hat{C}$ Eqs.(\ref{ptg},\ref{ptc}),
we get the resummed part of the anomalous dimension in the DIS scheme,
\be
 \hat{\gamma}^{DIS} = N \sum^{\infty}_{k=1}\hat{\gamma}^{k}
      \left( \frac{\bar{\alpha}_{s}}{N^{2}} \right)^{k}
  + \beta_{0} N^2 \sum^{\infty}_{k=2}
    \hat{d}^{k}
    \left( \frac{\bar{\alpha}_{s}}{N^{2}} \right)^{k} 
  + {\cal O} \left( N^3 
      \left( \frac{\bar{\alpha}_{s}}{N^{2}} \right)^{k} \right) \ ,
     \label{resumdis}
\ee
where the second terms come from the resummed coefficient function
and $\hat{d}^k$ are numerical numbers independent of $N$.
The above equation tells us that the resummed coefficient function
belongs to the NLL order corrections in the context of the resummation approach.
Then, one must include the NLL order anomalous
dimension, which has not yet been available,
to see the NLL effects.
%---------- Section 3 ----------------------------------------

\section{Numerical Analysis}

\vspace{1mm}
\noindent
Now we come to the numerical analyses to show how the final results
are sensitive to the choice of the input parton densities.
In conjunction with the claim in Ref.~\cite{bartels},
we also consider the effects from the resummed coefficient function
with the hope that the inclusion of it could shed some light
on the size of the NLL corrections.

Our starting point is the expression,
\be
 g_{1}(Q^{2},x) = \int_{ c-i \infty}^{c+i \infty} 
    \frac{d N}{2 \pi i} x^{- N} \exp
\left(-\int_{ \alpha_{s}(Q_{0}^{2})}^{\alpha_{s}(Q^{2})}
       \frac{d\alpha_{s}}{\beta} \gamma^{DIS} \right)
           g_{1}(Q_{0}^{2},N) \ .
\label{eqn:g1-DIS}
\ee
The anomalous dimension $\gamma^{DIS}$ which includes the resummation
of ${\ln}^n x$ terms is organized as follows,
\be
\gamma^{DIS}(N) = \bar{\alpha}_{s} \gamma^{1}(N)
   + \bar{\alpha}_{s}^{2} \gamma^{2}(N) + K (N,\alpha_{s})
  - \beta \frac{\partial}{\partial \alpha_{s}}
   \ln \left( 1 + \bar{\alpha}_{s} c^{1} + H(N,\alpha_{s}) \right) \ ,
\label{eqn:anoma}
\ee
where $\gamma^{1,2}$ and $c^{1}$ are respectively the exact anomalous
dimension and coefficient function at the one and two-loop fixed order
perturbation theory.
$K(N,\alpha_{s})$ ($H(N,\alpha_{s})$) is the resummed anomalous
dimension Eq.(\ref{ptg}) (Eq.(\ref{ptc})) with $k = 1,2$ ($k = 0,1$)
terms being subtracted.

It is to be noted here that the anomalous dimension at $N=1$
should vanish due to the conservation of the (non-singlet) axial
vector current. The perturbation theory guarantees this symmetry order by order
in the $\alpha_{s}$ expansion. However, the
resummation of the leading singularities in $N$ does not respect
this symmetry. Therefore, we need to restore this symmetry 
\lq\lq by hand\rq\rq . Our choice is~\cite{blvo},
\[  K(N,\alpha_{s}) \to K(N,\alpha_{s}) (1-N) \ .\]
Of course, this is not a unique prescription
\footnote{Our final conclusion remains the same qualitatively
if we choose other prescription.}.
We have used the technique in Ref.~\cite{Reya}
to perform the Mellin inversion Eq.(\ref{eqn:g1-DIS}).

We choose the starting value of the evolution to be
$Q_{0}^{2} = 4 GeV^{2}$. We calculate the $Q^2$ evolution for
two types of the input densities A and B:
A is a function which is flat at small $x$ ($x^{\alpha},\alpha \sim 0$),
and B is slightly steep ($\alpha \sim - 0.2$).
The explicit parametrization we use is~\cite{grsv},
\[ \Delta q (Q_0^{2},x) = N (\alpha , \beta , a )
       \eta x^{\alpha} (1 - x)^{\beta} ( 1 + a x)\ ,\]
where $N$ is a normalization factor such that
$\int dx N x^{\alpha} (1 - x)^{\beta} ( 1 + a x) = 1$
and $\eta = \frac{1}{6} g_A $ ($g_A = 1.26 $) in
accordance with the  Bjorken sum rule.
A and B correspond to the following values of parameters,
\[  A \ (B)\ :\  \alpha = + 0.0 \ (- 0.2)\  , \ \beta = 3.09 \ (3.15)\  , \
    a = 2.23 \ (2.72)\ .\]
We put the flavor number 
$n_{f} = 4 $ and $\Lambda_{QCD}=0.23 GeV$.

First we estimate the case which includes only the LL correction
$\hat{\gamma}$.
The evolution kernel in this case is obtained by dropping   
$H(N,\alpha_{s})$ in Eq.(\ref{eqn:anoma}).
Fig.1a (1b) shows the LL results (dashed curves) after evolving
to $Q^{2} = 10, 10^2 ,$ $10^4 GeV^{2}$ from the A (B) input density
(dot-dashed line). The solid curves are the predictions of
the NLO-DGLAP evolution. 
These results show a tiny enhancement compared with the NLO-DGLAP
analysis and are consistent with those in Ref.~\cite{blvo}
The enhancement is, as expected, bigger when the input density is
flatter. However any significant differences are not seen between the
results from different input densities.

Next, we include the NLL corrections from the resummed
\lq\lq coefficient function\rq\rq .
We show the results in Fig.2 by the dashed curves.
The results are rather surprising. The inclusion of
the coefficient function leads to a strong suppression
on the evolution of the structure function at small $x$.
Since the effects from the coefficient function fall in the NLL
level, the LL terms are
expected to (should) dominate at the small $x$. However our results
imply that the LL approximation
is not sensible in the small $x$ region we are interested in.
As the resummed coefficient function is only a part of the
NLL correction, we can not present a definite conclusion on the
(full) NLL correction. But it is obvious that the NLL correction
is very important in the experimentally accessible region of $x$.
%%%%%%%%%%% Fig.1 %%%%%%%%%%%%%%%%%%%%%%%%%%%%%%%%%%%%%%%%
\begin{figure}[H]
\begin{center}
\begin{tabular}{cc}
\leavevmode\psfig{file=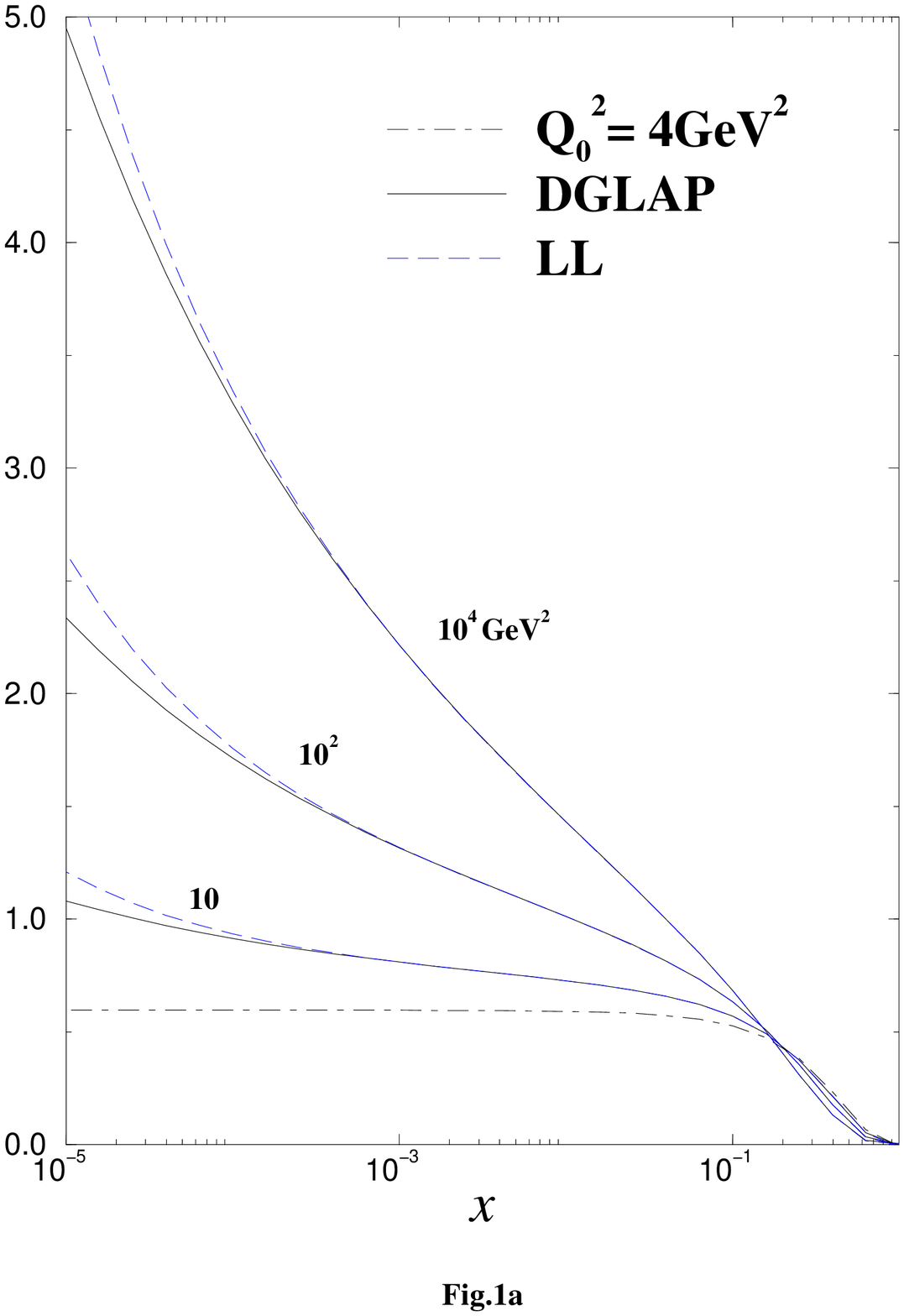,width=6.5cm} &
\leavevmode\psfig{file=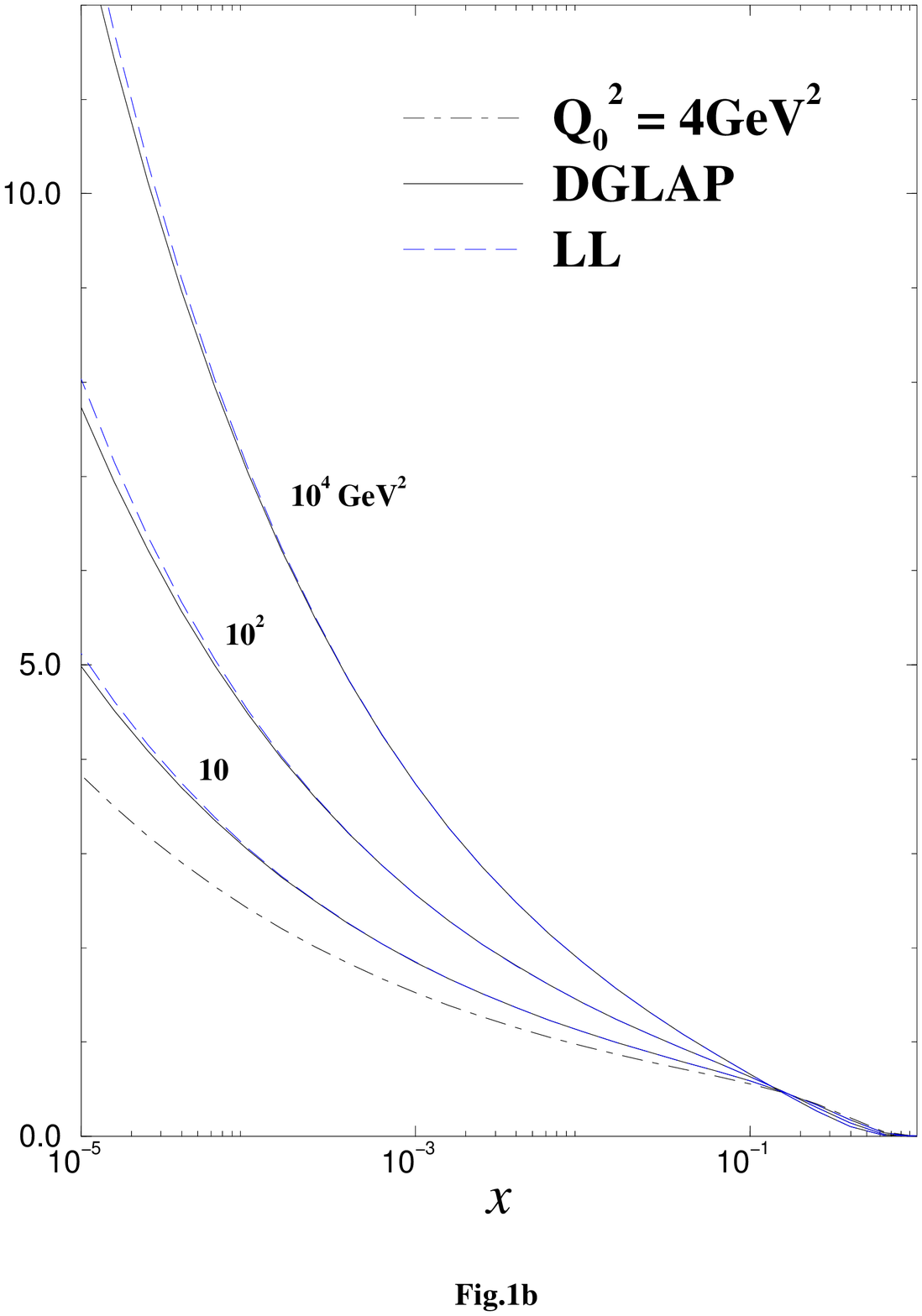,width=6.5cm} 
\end{tabular}
\caption{The LL evolution as compared to the DGLAP results with the flat
input A (1a) and steep one B (1b).}
\begin{tabular}{cc}
\leavevmode\psfig{file=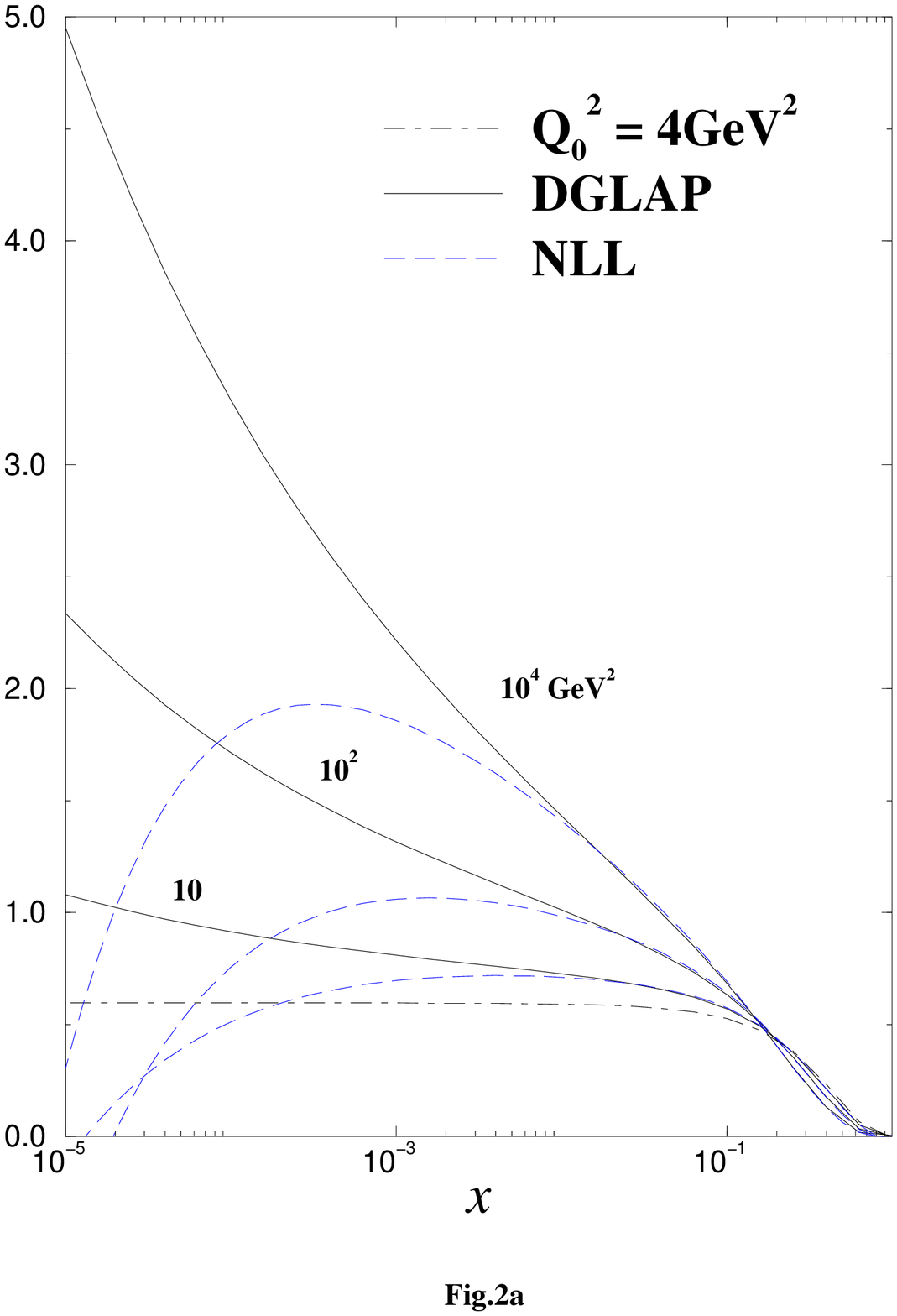,width=6.5cm} &
\leavevmode\psfig{file=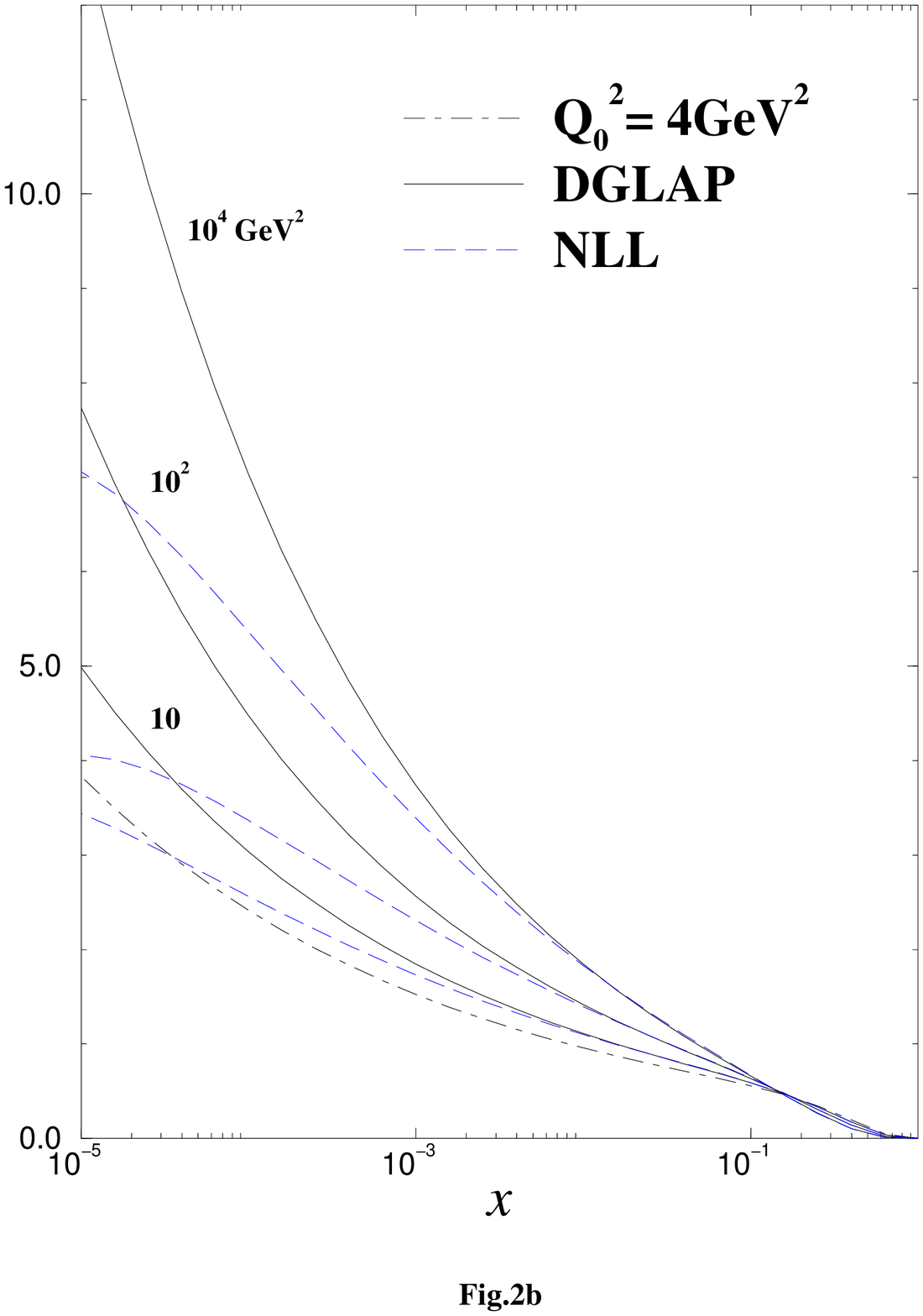,width=6.5cm} 
\end{tabular}
\caption{The NLL evolution as compared to the DGLAP results with the
flat input A (2a) and steep one B (2b). }
\end{center}
\end{figure}
%%%%%%%%%%%%%%%%%%%%%%%%%%%%%%%%%%%%%%%%%%%%%%%%%%%
%-------------- discussion and summary ----------------
\section{Discussion}

\vspace{1mm}
\noindent
In the previous section, we have shown that although the LL resummed
effect is very small in the experimentally accessible region of $x$,
a part of the NLL resummed contribution from the coefficient function
drastically changes the predictions.
To understand these numerical results, it will be helpful to remember the
perturbative expansion of the resummed results
Eqs.(\ref{ptg},\ref{ptc}).
Using the explicit values $N_C = 3 , C_F = 4/3$, we obtain
for the anomalous dimension in the DIS scheme
Eq.(\ref{resumdis}),
\bea
 \hat{\gamma}^{DIS} &=& N \left[ - 0.212
       \left( \frac{\alpha_{s}}{N^{2}} \right) \right. \nonumber\\
    & & \qquad - \, 
       0.068 \left. \left( \frac{\alpha_{s}}{N^{2}} \right)^{2} -
       0.017 \left( \frac{\alpha_{s}}{N^{2}} \right)^{3} -
       0.029 \left( \frac{\alpha_{s}}{N^{2}} \right)^{4} 
           + \cdots \right] \nonumber\\
    &+&  \, N^2 \left[
       0.141 \left( \frac{\alpha_{s}}{N^{2}} \right)^{2} +
       0.119 \left( \frac{\alpha_{s}}{N^{2}} \right)^{3} +
       0.069 \left( \frac{\alpha_{s}}{N^{2}} \right)^{4} 
           + \cdots \right] \label{numbergdis}\\ 
    &+&  \,\,\,  \cdots \ \ .\nonumber
\eea
Here note that: (1) the perturbative coefficients of the LL terms
(the first part of Eq.(\ref{numbergdis})) are negative
and those of the higher orders are rather small number.
This implies that the LL corrections push up
the structure function compared to the fixed-order DGLAP
evolution, but the deviations are expected to be small. 
(2) the perturbative ones from the NLL terms (the second part of
Eq.(\ref{numbergdis})), however, are positive and
somehow large compared with those of the LL terms.
This positivity of the NLL terms has the effect of
decreasing the structure function. This fact that
the coefficients with both sign appear in the anomalous dimension
should be contrasted with
the case of the unpolarized structure function~\cite{bfkl}.

Now it might be also helpful to {\sl assume} that the saddle-point dominates
the Mellin inversion Eq.(\ref{eqn:g1-DIS}). We have numerically estimated
the approximate position of the
saddle-point and found that the saddle-point stays around
$N_{\rm SP} \sim 0.31$ in the region of $x \sim 10^{-5}$ to $10^{-2}$.
By looking at the explicit values of the coefficients in
Eq.(\ref{numbergdis}), the position of the saddle-point seems to suggest that
the NLL terms can not be neglected. Since the coefficients from the higher
order terms are not so large numerically, it is also
expected that the terms  which
lead to sizable effects on the evolution may be only first few terms in the
perturbative series in the region of $x$ we are interested in.
We have checked that the inclusion of the first few terms in
Eq.(\ref{numbergdis}) already reproduces the results of section 3.
Fig.3a (3b) shows the numerical results of the
contribution from each terms of the NLL corrections in
Eq.(\ref{numbergdis}) at $Q^{2} = 10^2 GeV^{2}$ with the A (B)
type input density.
The solid (dot-dashed) line corresponds to the NLL (LL) result.
The long-dashed, dashed and dotted lines correspond
respectively to the case in which
the terms up to the order $\alpha_{s}^2$, $\alpha_{s}^3$, $\alpha_{s}^4$,
are kept in the NLL contributions. One can see that the
dotted line already coincides with the full NLL (solid) line.
These considerations could help us to understand why the NLL corrections
turns out to give large effects.
%%%%%%%%%%%%%%%%% Fig.3 %%%%%%%%%%%%%%%%
\begin{figure}[H]
\begin{center}
\begin{tabular}{cc}
\leavevmode\psfig{file=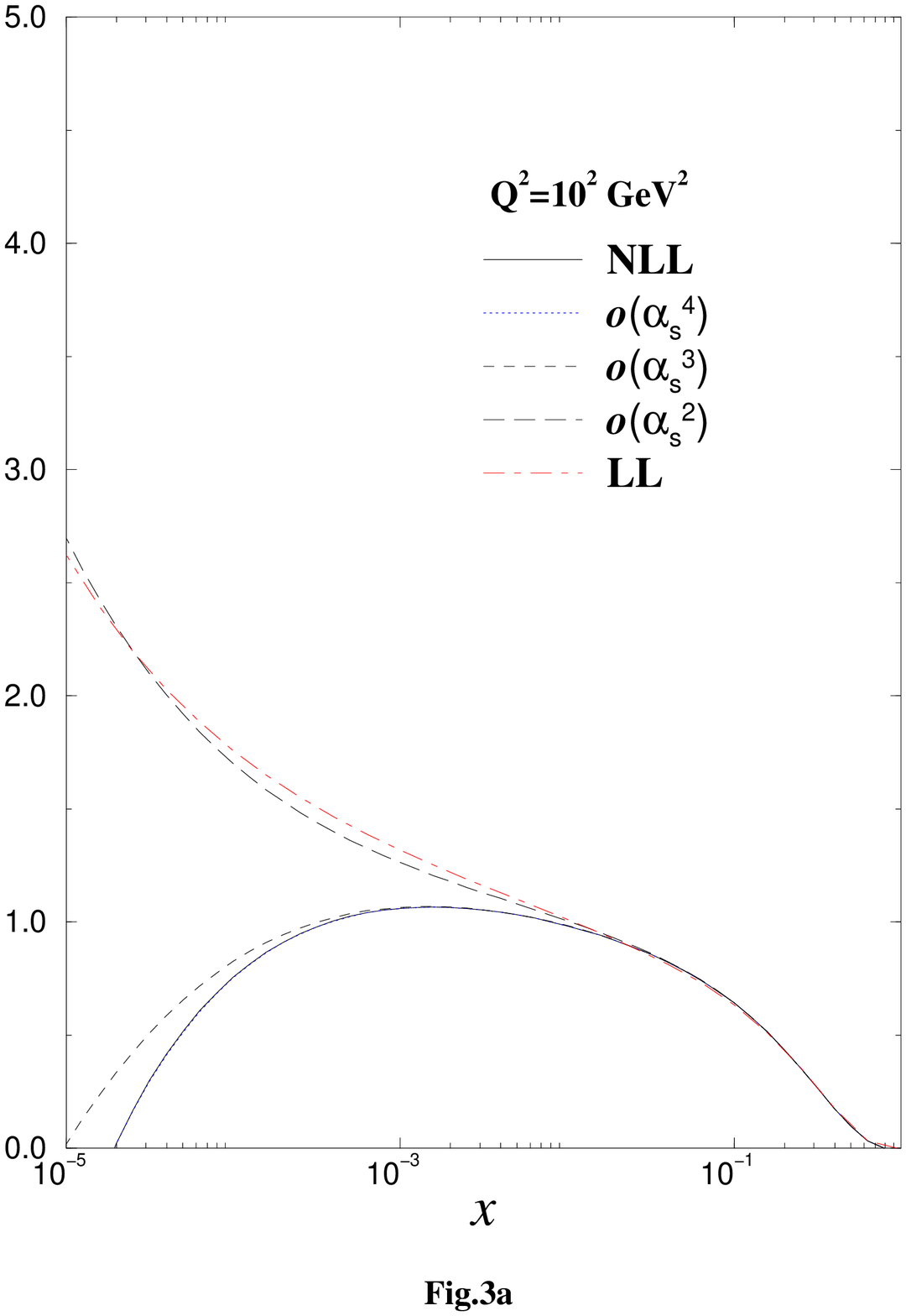,width=6.5cm} &
\leavevmode\psfig{file=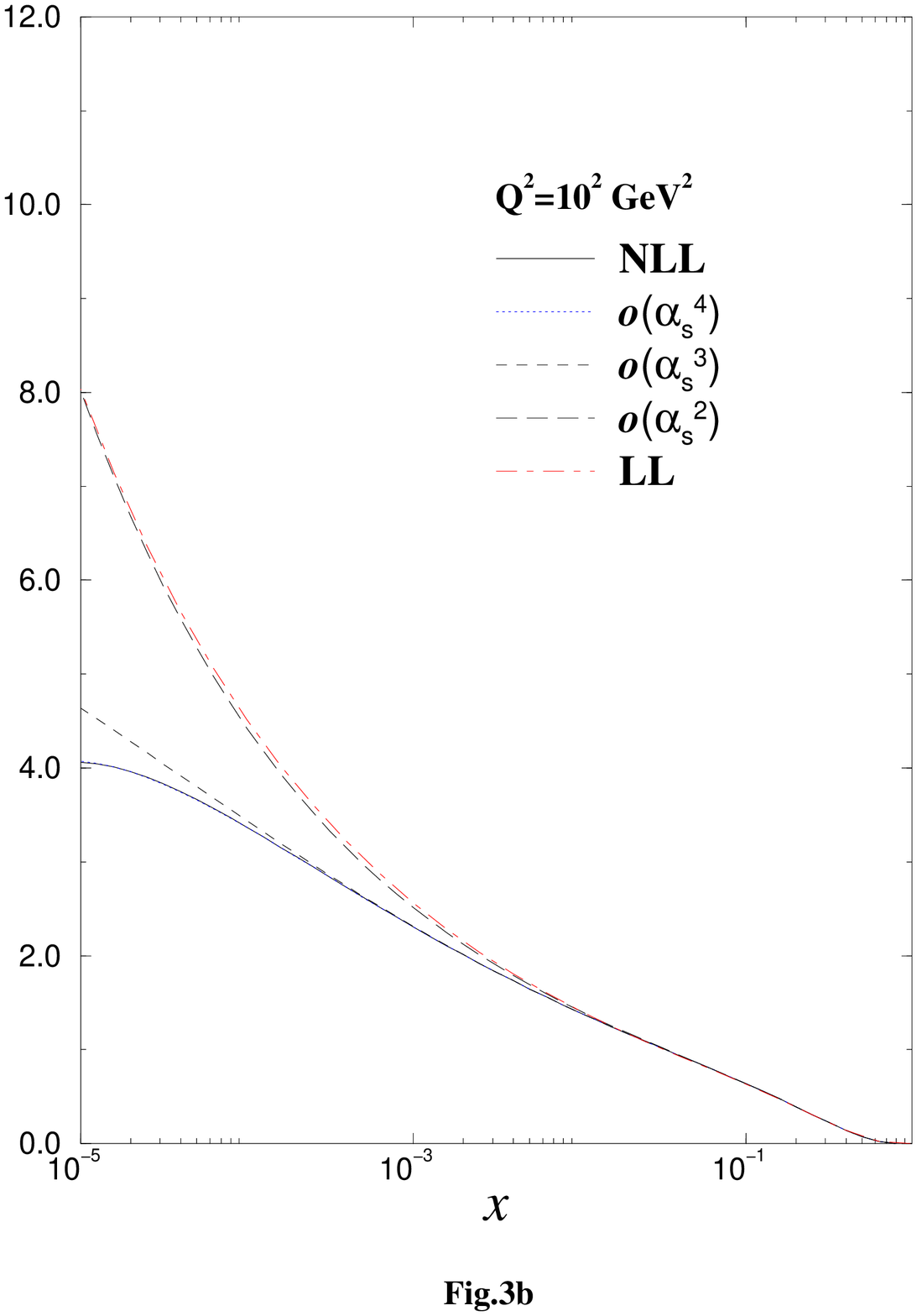,width=6.5cm} 
\end{tabular}
\caption{Contributions from the fixed order terms in the NLL
resummation with the flat input A (3a) and steep one B
(3b).}
\end{center}
\end{figure}
%%%%%%%%%%%%%%%%%%%%%%%%%%%%%%%%%%%%%
The final discussion concerns the convergence issue of the perturbative
series. As discussed in Refs.~\cite{bf}, one must be careful
when applying the perturbative approach to small $x$ evolution.
So according to Refs.~\cite{bf}, we have tried to solve
the evolution in $x$ space with first several terms
of the perturbative expansion being
kept and what we found is that the conclusion does not change.
The numerical results remain essentially the same.

In summary, we have performed numerical studies
for the flavor non-singlet $g_{1}$ at small $x$
by incorporating the all-order resummed anomalous dimension
and a part of the NLL corrections from the resummed coefficient
function. Including only the resummed coefficient
part is not theoretically consistent, and so one should take
into account also the anomalous dimension at the NLL level.
However, our results suggest that the LL analysis is unstable,
in the sense that a large suppression effect comes from the resummed
coefficient function which should be NLL correction.
We need a full NLL analysis to make a definite conclusion.

\vspace{1cm}
%------------------ Acknowledgement --------------------
\noindent
J.K. would like to thank the organisers of this
workshop for their hospitality.
This work was supported in part by the Monbusho
Grant-in-Aid Scientific Research No. A (1) 08304024,
No. C (1) 09640364 and the Monbusho International Scientific
Research No. 08044089.

%------------------ References -------------------------


\begin{thebibliography}{99}
%
\bibitem{intro}
   See {\sl e.g.} J. Bl\"umlein, S. Riemersma and A. Vogt, {\sl Preprint}
    DESY 96-131 / WUE-ITP-96-016 hep-ph/9608470; hep-ph/9610427
    and references therein.
\bibitem{bfkl}
   L. N. Lipatov, {\sl Sov. J. Nucl. Phys.} {\bf 23} (1976) 338;
   E. A. Kuraev, L. N. Lipatov and V. S. Fadin, {\sl Sov. Phys. JETP}
   {\bf 45} (1977) 199;
   Ya. Balitskii and L. N. Lipatov, {\sl Sov. J. Nucl. Phys.} {\bf 28}
   (1978) 822;
   S. Catani and F. Hautmann, {\sl Nucl. Phys.} {\bf B427} (1994) 475;
   S. Catani, {\sl Z. Phys.} {\bf C70} (1996) 263 and references therein.
\bibitem{smc}
   R. Windmolders, {\sl Talk at this Workshop};
   C. Young, {\sl Talk at this Workshop};
   D. Hasch, {\sl Talk at this Workshop.}
\bibitem{grsv}
   M. Stratmann, {\sl Talk at this Workshop.}
   S. Forte, {\sl Talk at this Workshop.}
%   G. Altarelli, R. D. Ball, S. Forte and G. Ridolfi, {\sl Nucl. Phys.}
%   {\bf B496} (1997) 337; 
%   M. Gl\"uck, E. Reya, M. Stratmann and W. Vogelsang, {\sl
%   Phys. Rev.} {\bf D53} (1996) 4775.
\bibitem{talk1}
   A. Vogt, {\sl Talk at this Workshop};
   R. Kirschner, {\sl Talk at this Workshop.}
\bibitem{bartels}
   J. Bartels, B. I. Ermolaev and M. G. Ryskin,
   {\sl Z. Phys.} {\bf C70} (1996) 273; {\sl ibid.} {\bf C72} (1997) 627.
\bibitem{kili}
   R. Kirschner and L. N. Lipatov, {\sl Nucl. Phys.} {\bf B213}
         (1983) 122.
\bibitem{blvo}
   J. Bl\"umlein and A. Vogt ,{\sl Phys. Lett.} {\bf B370}
   (1996) 149 ; {\sl Acta.Phys.Polonica} {\bf B27} (1996) 1309;
   J. Bl\"umlein, S. Riemersma and A. Vogt, hep-ph/9608470
\bibitem{kkt}
   Y. Kiyo, J. Kodaira and H. Tochimura, {\sl Z. Phys.}
   {\bf C74} (1997) 631.
\bibitem{koda}
   J. Kodaira, S. Matsuda, T. Muta, K. Sasaki and T. Uematsu,
      {\sl Phys. Rev.} {\bf D20} (1979) 627; 
   J. Kodaira, S. Matsuda, K. Sasaki and T. Uematsu,
      {\sl Nucl. Phys.} {\bf B159} (1979) 99.
\bibitem{van}
    R. Mertig and W. L. van Neerven, {\sl Z. Phys.} {\bf C70}
     (1996) 637 and references therein.
\bibitem{Reya}
   M. Gl\"uck, E. Reya and A. Vogt, {\sl Z. Phys.}
      {\bf C48} (1990) 471; 
   D. Graudenz, M. Hampel, A. Vogt and Ch. Berger, {\sl Z. Phys.}
      {\bf C70} (1996) 70.
\bibitem{bf}
   R. D. Ball and S. Forte, {\sl Phys. Lett.} {\bf B351} (1995) 313;
   J. R. Forshaw, R. G. Roberts and R. S. Thorne,
   {\sl Phys. Lett.} {\bf B356} (1995) 79.

\end{thebibliography}
\end{document}